\documentclass[acmsmall,screen]{acmart}

\AtBeginDocument{%
  }

\setcopyright{acmlicensed}
\copyrightyear{2026}
\acmYear{2026}

\acmConference[ASSETS '26]{The 28th International ACM SIGACCESS Conference on Computers and Accessibility}{October 26--29, 2026}{Denver, CO, USA}

\begin{document}

\title[When Sounds Hurt and Voices Aren't Heard]{When Sounds Hurt and Voices Aren't Heard: An Experience Report on Misophonia, Sensory Trauma, and Trauma-Informed Design}

\author{Tawfiq Ammari}
\email{tawfiq.ammari@rutgers.edu}
\orcid{0000-0002-1920-1625}
\affiliation{%
  \institution{School of Communication and Information, Rutgers University}
  \city{New Brunswick}
  \state{New Jersey}
  \country{USA}
}

\renewcommand{\shortauthors}{Ammari}

\begin{abstract}
This experience report reflects on researching misophonia as someone who lives with it. Misophonia is an aversive response to everyday sounds (chewing, pen clicking) and visual cues (misokinesia); it is poorly recognized, sufferers routinely disbelieved, and platform surfaces (auto-playing audio, algorithmic ASMR, eating on camera) turn the sensory environment into distress. I re-read 16 interviews with misophones through my lived experience and role in soQuiet Misophonia Research Network. I extend trauma-informed design (TID) in three ways: (1) TID must treat embodied, contested conditions as sources of sensory and epistemic harm; (2) closed groups and moderated subreddits reproduce dismissal when moderators decide whose experiences count, instituting personal-communal identity gaps in code; and (3) Randazzo and Ammari's Spinning Effects Model and grief bubbles describe how platform affordances erode misophones' disclosure efficacy and identity integration. Implications for ASSETS follow.
\end{abstract}



\maketitle

\section{Introduction and Positionality}

I write this report as someone who lives with misophonia and as a researcher who studies how design can support people whose experiences are routinely disbelieved. Misophonia is an aversive, often involuntary response to specific everyday sounds (chewing, slurping, sniffling, tongue clicking, pen clicking, keyboard typing, joint cracking) and, for many of us, to the visual cues that accompany them, a co-occurring sensitivity sometimes called \emph{misokinesia} \cite{jaswal2021misokinesia}. Recent consensus work defines misophonia as a distinct disorder of decreased tolerance to specific sounds \cite{swedo2022consensus}, but it has no DSM entry and no settled clinical home \cite{cavanna2015misophonia,schroder2013misophonia,taylor2017misophonia,potgieter2019misophonia}. People who live with it spend extraordinary effort just to be \emph{believed}, and, in parallel, extraordinary effort to make ordinary platform surfaces survivable.

I serve as a member of the soQuiet Misophonia Research Network \cite{soquiet_mrn,soquiet_ammari}, a community-led network that connects misophonia researchers to each other and to people who have lived experience of the condition. soQuiet's network was created in part because, for years, misophonia research happened \emph{to} people with misophonia rather than \emph{with} them; this asymmetry is common in disability research and is one ASSETS has long worked to interrupt \cite{spiel2020nothing,hofmann2020living,mack2021what,mankoff_disability_2010,bennett2019promise}. My positionality shapes the three contributions that follow.

\textbf{First}, I extend the conversation on trauma-informed design (TID) \cite{chen_trauma-informed_2022,Scott2023,randazzo_workshop_23,randazzo_ammari_23,randazzo_kintsugi_25} by foregrounding two harms that TID has tended to treat as separable. \emph{Sensory trauma} is the recurring, design-mediated harm of unavoidable exposure to audiovisual content one's body cannot tolerate. \emph{Epistemic trauma} is the cumulative harm of having one's first-person account of that exposure systematically disbelieved \cite{stinnett2018trauma,mohamed2024debilitating}. For misophones, these are not two stories but one: the sounds hurt, and the dismissal that follows hurts again. I position TID as an extension of Universal Design: the curb-cut logic applies, and sensory and recognition features that support misophones produce platforms more usable for everyone.

\textbf{Second}, drawing on a re-reading of 16 semi-structured interviews with people who live with misophonia \cite{ammari2026remote}, I show that the platforms misophones turn to for validation (closed Facebook groups, moderated subreddits, gated Discord servers) can themselves reproduce epistemic dismissal when a few moderators decide whose experiences count. This complicates the finding that closed, moderated communities are simply ``safer'' alternatives to open ones \cite{ammari_et_al_19,Andalibi_19}: the same gating that screened out hostile users also screened out members whose lived experience contradicted the moderator's theory.

\textbf{Third}, I reflect on what it has been like, as a researcher, to interview people about an experience I share. The interviews surfaced not only my participants' epistemic wounds but also some of my own.

\section{Background: Trauma-Informed Design Meets Sensory Trauma}
\label{sec:background}

\subsection{Trauma-Informed Design in HCI}

Trauma-informed approaches in HCI build on the SAMHSA framework's emphasis on safety, trustworthiness, peer support, collaboration, empowerment, and attention to cultural and historical context. Chen et al.~\cite{chen_trauma-informed_2022} brought these into computing as \emph{trauma-informed computing}, arguing that designers should anticipate that users may carry trauma and design systems that minimize re-traumatization. Scott et al.~\cite{Scott2023} extended the lens to social media, presenting the six SAMHSA principles as ``sensitizing concepts'' and including the \emph{physical} environment in which trauma occurs. Randazzo et al.~\cite{randazzo_workshop_23} push the field further by treating TID as fundamentally collaborative: indicators, metrics, and guidelines should be developed \emph{with} affected users, clinicians, and practitioners. Randazzo and Ammari develop this empirically. Their \emph{Spinning Effects Model}~\cite{randazzo_ammari_23} traces how platform affordances stabilize or destabilize a survivor's disclosure efficacy across three effects (\emph{activation}, when a narrative mirrors one's experience; \emph{toppling}, when affordances such as direct feedback or content persistence weaken confidence to speak; and \emph{steadying}, when anonymity, indirect feedback, or community recognition stabilize it), and they show that algorithmic filter bubbles can \emph{counteract} societal filter bubbles for ``muted'' users who lack settled language for their experience. In follow-on work, they introduce \emph{grief bubbles}~\cite{randazzo_kintsugi_25}: algorithmic recommendation systems that concentrate on the traumatic aspects of a user's identity and impede integration with the rest of who they are. Related work has examined content warnings, opt-in disclosure, and supportive moderation as ways to soften the platform's impact on users carrying difficult histories \cite{andalibi_understanding_2016,andalibi_self-disclosure_2017,Andalibi_19,andalibi_responding_2018}.

It is useful to position TID as an under-theorized extension of Universal Design. Where Universal Design removes physical and sensory barriers to participation, TID addresses the affective and cognitive conditions that determine whether people can engage at all. The \emph{curb-cut effect} is instructive: sidewalk curb cuts were designed for wheelchair users but came to benefit parents with strollers, delivery workers, cyclists, and travelers with luggage \cite{mankoff_disability_2010}. TID has the same logic: design choices that reduce re-traumatization, build in user control, and lower the cognitive cost of participation help not only trauma survivors but everyone. This report argues that the curb-cut benefits of TID have not yet been claimed for sensory-sensitive users, and that, when they are, they will be felt well beyond that population.

What the existing TID literature shares, however, is a focus on the \emph{content} flowing through platforms and on whether users have controls to manage exposure to it. Chen et al.~\cite{chen_trauma-informed_2022} note that trauma-informed safety includes physical conditions like ``keeping noise levels low''; Scott et al.~\cite{Scott2023} note that physical safety includes wall colors, lighting, and ``being attentive to signs of discomfort.'' In both, the sensory environment is framed as a relatively static stage on which content-mediated harms happen, rather than as itself the harm. For embodied and contested conditions like misophonia, that framing under-specifies what trauma-informed design actually has to do.

\subsection{The Sensory Surface as a Site of Ongoing Trauma}
\label{sec:sensory-surface}

My prior work on misophonia and technology non-use \cite{ammari2026remote} surfaces audiovisual challenges that are not edge cases but constitutive of how mainstream platforms work. TikTok's For-You algorithm surfaces ASMR (deliberately triggering tapping, scratching, whispering) as ``relaxing'' content; Instagram and YouTube auto-play sounds; Zoom and Discord present a single shared audio mix in which a colleague's chewing or pen-clicking sits at the same volume as their speech; and eating on camera, atypical in person, has become a normal video-meeting behavior. For someone with misophonia, these are not background design choices: they are recurring exposures to material the body responds to with anger, panic, or fight-or-flight \cite{swedo2022consensus,jaswal2021misokinesia}. Visual triggers (misokinesia) compound this; muting the audio does not silence the sight of someone chewing or fidgeting on screen.

Read through Randazzo and Ammari's lens, the For-You page that keeps serving ASMR to a misophone is a \emph{sensory grief bubble}~\cite{randazzo_kintsugi_25}: the algorithm has learned that this user engages with content about misophonia and responds by surfacing more of the very stimuli that triggered the engagement. Treating trauma only as carried-in history misses that, for sensory-sensitive users, the platform's audiovisual surface is itself the re-traumatizing event: each unmuted call, each algorithmically surfaced ASMR clip, each unannounced mukbang\footnote{A video in which a host consumes large amounts of food while interacting with their audience.} is harm produced by the system, not merely passing through it. Earlier accessibility work \cite{race_et_al_21,zolyomi2021social,poulsen2024auditory,das_et_al_21} and recent neurodivergent-design scholarship \cite{jiang2025shifting,sabinson2024pictorial,sabinson2025blowfish,baillargeon2025social,li2024codesigning,nevskyetal25} chart the relevant vocabulary. Reading TID through this material reframes Chen et al.'s ``noise levels'' and Scott et al.'s ``calming colors'' as a core obligation: a trauma-informed platform for sensory-sensitive users would treat the audiovisual surface as a moderation domain on the same footing as content, with design strategies (disaggregated audiovisual control, real-time sensory filtering, sensory predictability, collaborative boundary negotiation, and visual trigger management \cite{ammari2026remote}) available by default rather than negotiated case by case.

\subsection{Epistemic Injustice and Epistemic Trauma}

Sensory harm is half the story. The other half is who is believed when they describe it. Fricker~\cite{fricker_epistemic_2007} introduced \emph{epistemic injustice} to name a wrong done to someone in their capacity as a knower. Two forms matter here: \emph{testimonial injustice}, where a speaker is given less credibility than they are due, and \emph{hermeneutical injustice}, where no shared interpretive framework is available for someone to make sense of their experience. Carel and Kidd~\cite{carel2014epistemic,carel2017epistemic,kidd2017epistemic} developed this lens for healthcare; Newbigging and Ridley~\cite{newbigging2018epistemic} and McKinnon~\cite{mckinnon2016epistemic} document the pattern in mental health; Bhakuni and Abimbola~\cite{bhakuni_epistemic_2021} extend it into global health. Misophonia is a particularly clear case: no DSM entry, no consensus biomarker, only a recent published consensus definition \cite{swedo2022consensus}. Misophones face hermeneutical injustice (no widely shared vocabulary) and testimonial injustice (accounts doubted or pathologized) at once. The cumulative effect is what Stinnett~\cite{stinnett2018trauma} names \emph{epistemic trauma}: ``after so [many] repeated instances of silencing, she begins to doubt her own experiences\ldots and eventually no longer feels comfortable sharing'' \cite[p.~45--46]{stinnett2018trauma}. Mohamed~\cite{mohamed2024debilitating} notes this trauma can be transmitted, affecting both the person sharing and the professional, or the researcher, receiving.

For HCI, the consequence is that two TID questions need to be asked together: \emph{whose body is the platform's audiovisual surface designed for?}, and \emph{whose account of that surface is the platform structured to take seriously?} ASSETS already foregrounds the structural version of the second: Spiel et al.~\cite{spiel2020nothing} insist that ``nothing about us'' should be done ``without us''; Hofmann et al.~\cite{hofmann2020living} and Mack et al.~\cite{mack2021what} interrogate which lived experiences accessibility research is built on; Bennett and Rosner~\cite{bennett2019promise} argue that empathy without lived expertise can mask whose authority is being centered.

\section{Approach}
\label{sec:methods}

This experience report is a re-reading of an earlier paper \cite{ammari2026remote}, a qualitative analysis of 16 semi-structured interviews with U.S.-based adults who live with misophonia, recruited through online misophonia communities. The full study design, participant table, recruitment, and analytic procedure are reported there. Here I draw on three intertwined sources: my own lived experience with misophonia, my work with the soQuiet Misophonia Research Network \cite{soquiet_mrn}, and a return to those 16 transcripts with two specific lenses. The first lens asks where participants described specific platform surfaces (auto-playing audio, ASMR feeds, video-call audio mixes, eating on camera) producing distress that platform features could not let them manage. The second asks where they described being unable to share that experience, or described specific platform structures (moderation, gated access, single-moderator authority) that made it harder rather than easier to be believed. I read those passages alongside the literature on epistemic injustice \cite{fricker_epistemic_2007,carel2014epistemic,stinnett2018trauma,mohamed2024debilitating} and on trauma-informed design \cite{chen_trauma-informed_2022,Scott2023,randazzo_workshop_23,randazzo_ammari_23,randazzo_kintsugi_25}. This document does not report new findings; it is a reflective re-reading, consistent with prior ASSETS work that treats researcher positionality as an analytic resource rather than a confound \cite{hofmann2020living,mack2021what,branham2015invisible}.

I share many of the triggers participants described. I have had clinicians look at me blankly when I named the condition, and I have left online communities I once relied on because of similar tensions. Working with the soQuiet network has made plain to me that ``misophonia community'' is not a monolith: it is a fragile, internally contested space, and the boundaries of who is a credible speaker inside it are themselves under negotiation.

\section{What Participants Could Not Manage, and What They Could Not Share}
\label{sec:findings}

Across the 16 interviews, four patterns stood out. The first concerns platform surfaces participants could not turn off, mute, or filter without abandoning the platform entirely; the second and third turn on whose account is treated as legitimate; the fourth on how the platform's identity and feedback affordances shape who speaks at all.

\subsection{Sensory Trauma in Platform Surfaces}
\label{sec:findings-sensory}

Participants described the audiovisual fabric of mainstream platforms as a site of recurring distress, not background. Kathleen and Cindy each described TikTok as a platform whose value proposition (audio-synchronized content delivered without sensory pre-screening) put them in an impossible position: Kathleen had a ``love-hate'' relationship with the platform because the algorithm kept surfacing ASMR she could not tolerate, while Cindy left the platform altogether because she could not watch the videos without sound and watching with sound meant constant exposure \cite{ammari2026remote}. Read against TID, neither account is a content-moderation problem in the standard sense: the sounds involved are ASMR clips and mukbangs delivered as ``relaxing'' content.

Video conferencing produced a related pattern. Rhys characterized Zoom as ``so auditory'' that long sessions became sustained trigger exposure she could only blunt with coarse mute-all/hear-all controls. Nathan described eating on camera as both an audio and a visual trigger: ``I'm not fond of people eating on Zoom\ldots I don't want to hear it. And I don't want to watch it\ldots if hearing them is maybe a nine, you know, seeing it as, like a two or three.'' Amelia reported confronting a music teacher who ate during lessons, ``feel[ing] weird telling her this'' because the disclosure made her, not the eating, the social problem. Participants consistently described the binary mute/unmute structure as itself the issue: no per-participant volume, no separation of speech from ambient sound, no way to blur a single video feed without affecting the rest \cite{ammari2026remote}.

What makes these accounts traumatizing is their unannounced, unfilterable, recurring delivery into bodies that respond with fight-or-flight. The Race et al.~\cite{race_et_al_21} auditory design recommendations and the Jiang et al.~\cite{jiang2025shifting} ADHD-inclusive principles point at the right vocabulary. What is needed at the platform level is to treat that vocabulary as part of trauma-informed design's ``safety'' principle, not as a separate accessibility silo.

\subsection{``Are You Lying?'': Disbelief in Clinical and Family Settings}

Participants repeatedly described being disbelieved when they tried to name their experience. Evelyn told us she felt she had to come to interactions armed with research articles in order to be taken seriously; the everyday disbelief she encountered (``Are you lying?'') made her feel ``unimportant'' \cite{ammari2026remote}. That phrasing is an exact description of testimonial injustice \cite{fricker_epistemic_2007}: not that one's testimony is rebutted on the merits, but that one is not granted standing as a knower of one's own life.

Kathleen, who has a master's in psychology and works as an organizational psychologist, described seeing three or four specialists and several general clinicians without finding anyone who could help. Substantial professional literacy does not protect against the structural absence of credible interlocutors \cite{ammari2026remote}. Lydia, looking back on years before she had a name for what she was experiencing, described what hermeneutical injustice feels like from inside: ``I, like, pretty much thought I was insane from\ldots the whole time, really.'' Lydia and others also described family members who, after disclosure, made the trigger sounds on purpose; a relative would open a bag of chips with ``oh, is this gonna be too loud for you?'' Disclosure followed by deliberate triggering is a specific weaponization of the epistemic position the discloser was trying to claim. The harm is not only that recognition fails; the discloser is repeatedly required to \emph{prove themselves} against the default assumption that they are exaggerating.

\subsection{When the Online Community Becomes a Gated Community}

For people who do not get recognition from clinicians or family, online community is often the first place they feel believed. But the platforms participants used to reach those communities are not neutral: they impose structures of access that decide whose accounts get heard.

Several participants described private, moderator-gated groups in mixed terms: the same gating that made these spaces feel safer also kept members from finding them, reading them, or speaking inside them. Lola had joined a private Facebook misophonia group years earlier and almost forgotten about it. The barrier was partly Facebook's affordances (the platform was ``not as accessible'' to her) and partly the learned expectation, after years of dismissal, that sharing would go badly. Justine described herself as ``too much of a private person\ldots Sometimes I write [a comment] down\ldots and then I'll delete it.'' Chuck named the moderation gating trade-off directly: gated subreddits screen out trolls, but they also screen out the very people who might most benefit—those unwilling to apply to a group whose contents they cannot see in advance \cite{ammari2026remote}.

Of all sixteen interviews, Ashley's most directly informed the framing I adopt here. She had joined a closed Facebook group run by a single physician-moderator and initially valued it. When she brought in research arguing that misophonia has psychological and neurological dimensions alongside the auditory (a position consistent with the broader literature \cite{swedo2022consensus,cavanna2015misophonia,potgieter2019misophonia,jaswal2021misokinesia,schroder2013misophonia,taylor2017misophonia}) and named her own visual triggers as embodied evidence, the moderator rejected the psychological framing outright and banned her, in Ashley's words, ``just for asserting that.''

Ashley was not removed for harassment. She was removed for using her own embodied evidence to push back on the moderator's theory. The platform feature that gave the moderator authority to keep harassers out also gave her authority to keep contradictory lived experience out. Read through Randazzo and Ammari's lens~\cite{randazzo_kintsugi_25}, what Ashley encountered was a \emph{personal-communal identity gap}: a discrepancy between her own embodied understanding of misophonia (her personal frame) and how the community's moderation infrastructure labeled and bounded that experience (the communal frame she was expected to inhabit). The gap was not a mismatch she could close by speaking more clearly; it was a gap the platform's moderation architecture was actively maintaining. From inside the group the moderator's theory looked like settled fact; from outside, a single gatekeeper had used the platform's moderation affordances to enforce her interpretive frame against members whose first-person accounts did not fit. This is the structural shape of testimonial injustice, implemented in code.

Ashley described a second harm: harassment that followed her across subreddits. After a disagreement on an unrelated subreddit, another user pulled her misophonia posts forward and used them against her, ``denigrating me for, you know, this misophonia that I have and like, you're crazy.'' She reported the user to multiple moderation teams, including the misophonia subreddit's. ``Nobody did anything. Nobody cared at all,'' she said. The first incident showed her what concentrated moderator authority does when it goes wrong; the second showed her what diffuse moderator authority does when it does nothing at all.

\subsection{Anonymity, Asymmetric Feedback, and Self-Censorship}

A quieter pattern runs through the transcripts: people whose willingness to share is shaped by the platform's identity and feedback affordances. Several participants named anonymity as the precondition for sharing at all. Kathleen described ``a lot of shame'' around the condition and explained why Reddit was the only space she used for it: ``Reddit is truly the one place where it's completely anonymous\ldots that feels very safe.'' Vlad framed the calculus more sharply: in non-anonymous spaces, ``best case scenario, people are not going to be able to relate, and worst case scenario, they'll be dismissive, or even put it on you.'' Lola, who has an Instagram account, will not post about misophonia by name; the gate is identifiability rather than reach. What Reddit's pseudonymous architecture supplies for these participants is not anonymity-as-privacy in the usual sense, but \emph{anonymity-as-credibility-protection}: a way to speak without having one's standing on the topic adjudicated against an offline identity. In Randazzo and Ammari's terms~\cite{randazzo_ammari_23}, this is a \emph{steadying} effect, where pseudonymity stabilizes confidence to disclose for users who would otherwise be muted by repeated dismissal.

Reddit's voting affordances cut the other way. Kendra valued misophonia subreddits as ``a realm that's not just going to downvote me into oblivion''; Kathleen, who replies to younger users' posts, observed that ``people don't really reply back that often\ldots they just kind of like upvote the posts.'' This is what Randazzo and Ammari call \emph{indirect feedback}~\cite{randazzo_ammari_23}: users assimilate the votes and replies received by others' posts as if directed at themselves, internalizing what is sayable in the group before they ever post. The platform's primary feedback channel is a binary signal (acceptable or not), not a substantive sign that the experience behind the post has been recognized; for misophones already trained to expect dismissal, a single visible downvote can do the same work as an in-person eye-roll.

A third response is outright withholding. Justine writes comments and deletes them. Evelyn ``mostly go[es] on there just to like, be validated\ldots I don't make any posts myself.'' Kathleen reads everything and posts almost nothing of her own. Many participants combined active community membership with very low public visibility: they read, they upvoted, they took screenshots, but they did not author. Randazzo and Ammari~\cite{randazzo_ammari_23} reframe this kind of lurking as a strategic self-care behavior rather than freeloading: a defense mechanism for users whose disclosure efficacy has been repeatedly destabilized. In Stinnett's~\cite{stinnett2018trauma} language, repeated silencing leads people to doubt their own experiences and ``eventually no longer feel\ldots comfortable sharing.'' Self-censorship is not a personality trait; it is what epistemic trauma looks like as a long-run user-behavior pattern, shaped by the identity and feedback affordances the platform makes available.

\subsection{Widened or Glazed: Technological Affordances and Trauma-Informed Design}

The original paper \cite{ammari2026remote} read these transcripts through platform affordances and selective domestication \cite{salovaara2011everyday,sorensen2006domestication,gorm2019episodic,baumer_et_al_13,baumer_et_al_14}, producing a five-principle design framework that still holds. What it underweighted is how cleanly those principles sit inside trauma-informed design, and how directly the patterns map onto Randazzo and Ammari's models~\cite{randazzo_ammari_23,randazzo_kintsugi_25}. Three reframings emerge. \textbf{(1)~Sensory accessibility is trauma-informed safety:} ASMR auto-play and unfilterable Zoom audio are TID violations, not edge-case accessibility issues. \textbf{(2)~Recognition is a platform feature:} whether a user is believed when they describe their condition is shaped by who is granted moderation authority and what evidence counts. Closed, moderated communities, often valorized as ``safer'' \cite{ammari_et_al_19,Andalibi_19}, can institutionalize a single moderator's theory. \textbf{(3)~Misophonia produces fractured identities that platforms widen or glaze.} Ashley's clash with the moderator, Lydia's years of wondering whether she was ``insane,'' Kathleen's shame, and Evelyn's habit of arriving at every conversation armed with research articles are instances of the identity fractures Randazzo and Ammari~\cite{randazzo_kintsugi_25} describe; whether those fractures widen or are glazed is a function of platform design, not personal resilience.

\section{Reflection: Researching What You Live With}
\label{sec:reflection}

When Ashley described being kicked out of a group for naming her visual triggers, I recognized her account immediately, because I have visual triggers, and because I have argued for the legitimacy of those triggers in spaces where the dominant theory of misophonia treats it as purely auditory. When Kathleen described seeing four specialists and finding none who could help, I recognized that too. The interviews were specifically painful in the sense Mohamed~\cite{mohamed2024debilitating} describes: epistemic trauma can pass between the person sharing and the person receiving. The same is true on the sensory side. I was the researcher, but I was also the person who has had to stop a Zoom call because of someone else's lunch.

Two commitments followed. On disclosure: I named my misophonia when relevant but tried not to use it as a credentialing move; I wrote detailed memos after each interview, treating my reactions as data rather than background, consistent with autoethnographic and autobiographical-design traditions in HCI \cite{branham2015invisible,sabinson2025blowfish}. On accountability: my affiliation with the soQuiet Misophonia Research Network \cite{soquiet_mrn} means the people whose experiences I write about are people I will see at workshops and on calls. The arguments here have to be defensible to the people they describe. ASSETS has long been a venue where this kind of accountability is taken seriously \cite{spiel2020nothing,hofmann2020living,mack2021what,bennett2019promise}.

\section{Implications for the ASSETS Community}
\label{sec:implications}

\paragraph{1.~Make sensory and epistemic harm jointly visible inside trauma-informed design.}
TID conversations \cite{chen_trauma-informed_2022,Scott2023,randazzo_workshop_23,randazzo_ammari_23,randazzo_kintsugi_25} have largely treated trauma as carried-in history that platforms should not reactivate, with the sensory environment as a static stage. For embodied and contested conditions, trauma is also being produced \emph{in real time} by the audiovisual surface and by interactions that doubt the person's account of their own body. Designs that take this seriously should ask three questions together: does this surface re-expose users to triggering content; does it let sensory-sensitive users participate without an all-or-nothing choice between presence and physiological distress; and does it treat users as credible knowers of their own condition? Concretely: treat the five strategies from \cite{ammari2026remote} (disaggregated audiovisual control, real-time sensory filtering, sensory predictability, collaborative boundary negotiation, and visual trigger management) as TID safety features; pair them with moderation policies that name first-person accounts as a recognized form of evidence, anonymity-preserving identity affordances, and recognition mechanisms (badges, peer endorsements, substantive replies rather than only votes) that distribute credibility across community members. By curb-cut logic, these are not niche accommodations: per-participant volume, ASMR-aware feed controls, and credibility-distributing recognition help misophones first, but they help anyone whose lived experience is contested.

\paragraph{2.~Treat moderation architecture as an accessibility feature.}
ASSETS researchers already treat captioning, alt text, and contrast as accessibility features. The data suggest that moderation architecture (who can ban, who can pin, whose post is removed and on what grounds) is also an accessibility feature for people whose condition is contested. Concentrated moderator authority can institutionalize a single theory of the condition at the cost of members whose embodied experience contradicts it. Distributed approaches, such as rotating moderators, multi-moderator consensus for membership decisions, transparent appeals processes, and a separation of harassment-removal from interpretive-disagreement-removal, are accessibility infrastructure, not just governance \cite{Scott2023,andalibi_self-disclosure_2017,Andalibi_19,ammari_et_al_19,randazzo_workshop_23}. Randazzo and Ammari's proposal for a public \emph{moderation transparency metric}~\cite{randazzo_kintsugi_25}, visible to prospective members before they join, is a concrete first step.

\paragraph{3.~Make space for researchers with the conditions they study.}
The accessibility research community has come a long way on lived expertise \cite{spiel2020nothing,hofmann2020living,mack2021what,bennett2019promise}. What still needs more explicit support is the labor of researchers who carry the conditions they study into the research itself, including emotional and sensory labor that does not fit into project budgets. Concrete steps include protecting and celebrating autoethnographic and experience-report tracks; funding peer accountability structures like soQuiet's research network \cite{soquiet_mrn}; and treating researcher positionality not as confessional disclosure at the start of a paper but as an analytic resource that runs through it.

\section{Limitations and Conclusion}
\label{sec:conclusion}

This report is not a generalizable empirical study. The 16 interviews are a small, U.S.-based, predominantly white and female sample; participants who never found those communities are by construction underrepresented, and another researcher could surface different patterns. Future participatory work led by misophones should test, complicate, and where necessary contradict what I have argued. For people who live with misophonia, the harm is not only the sounds; it is the dismissal that follows. Trauma-informed design must take the audiovisual surface seriously as a site of \emph{sensory} trauma and moderation architecture as a site of \emph{epistemic} trauma~\cite{chen_trauma-informed_2022,Scott2023,randazzo_workshop_23,randazzo_ammari_23,randazzo_kintsugi_25}, and ASSETS must keep making room for researchers who write from inside the conditions they study.

\bibliographystyle{ACM-Reference-Format}
\bibliography{sample-base}


\begin{thebibliography}{49}


\ifx \showCODEN    \undefined \def \showCODEN     #1{\unskip}     \fi
\ifx \showISBNx    \undefined \def \showISBNx     #1{\unskip}     \fi
\ifx \showISBNxiii \undefined \def \showISBNxiii  #1{\unskip}     \fi
\ifx \showISSN     \undefined \def \showISSN      #1{\unskip}     \fi
\ifx \showLCCN     \undefined \def \showLCCN      #1{\unskip}     \fi
\ifx \shownote     \undefined \def \shownote      #1{#1}          \fi
\ifx \showarticletitle \undefined \def \showarticletitle #1{#1}   \fi
\ifx \showURL      \undefined \def \showURL       {\relax}        \fi
\providecommand\bibfield[2]{#2}
\providecommand\bibinfo[2]{#2}
\providecommand\natexlab[1]{#1}
\providecommand\showeprint[2][]{arXiv:#2}

\bibitem[Ammari and Gilgan(2026)]%
        {ammari2026remote}
\bibfield{author}{\bibinfo{person}{Tawfiq Ammari} {and} \bibinfo{person}{Samantha Gilgan}.} \bibinfo{year}{2026}\natexlab{}.
\newblock \showarticletitle{Remote Triggers: Misophonia, Technology Non-Use, and Design for Inclusive Digital Spaces}.
\newblock \bibinfo{journal}{\emph{arXiv preprint arXiv:2601.13355}} (\bibinfo{year}{2026}).
\newblock


\bibitem[Ammari et~al\mbox{.}(2019)]%
        {ammari_et_al_19}
\bibfield{author}{\bibinfo{person}{Tawfiq Ammari}, \bibinfo{person}{Sarita Schoenebeck}, {and} \bibinfo{person}{Daniel Romero}.} \bibinfo{year}{2019}\natexlab{}.
\newblock \showarticletitle{Self-Declared Throwaway Accounts on Reddit: How Platform Affordances and Shared Norms Enable Parenting Disclosure and Support}.
\newblock \bibinfo{journal}{\emph{Proc. ACM Hum.-Comput. Interact.}} \bibinfo{volume}{3}, \bibinfo{number}{CSCW}, Article \bibinfo{articleno}{135} (\bibinfo{date}{nov} \bibinfo{year}{2019}), \bibinfo{numpages}{30}~pages.
\newblock
\href{https://doi.org/10.1145/3359237}{doi:\nolinkurl{10.1145/3359237}}


\bibitem[Andalibi(2017)]%
        {andalibi_self-disclosure_2017}
\bibfield{author}{\bibinfo{person}{Nazanin Andalibi}.} \bibinfo{year}{2017}\natexlab{}.
\newblock \showarticletitle{Self-disclosure and {Response} {Behaviors} in {Socially} {Stigmatized} {Contexts} on {Social} {Media}: {The} {Case} of {Miscarriage}}. In \bibinfo{booktitle}{\emph{Proceedings of the 2017 {CHI} {Conference} {Extended} {Abstracts} on {Human} {Factors} in {Computing} {Systems}}} \emph{(\bibinfo{series}{{CHI} {EA} '17})}. \bibinfo{publisher}{ACM}, \bibinfo{address}{New York, NY, USA}, \bibinfo{pages}{248--253}.
\newblock
\showISBNx{978-1-4503-4656-6}
\href{https://doi.org/10.1145/3027063.3027137}{doi:\nolinkurl{10.1145/3027063.3027137}}


\bibitem[Andalibi(2019)]%
        {Andalibi_19}
\bibfield{author}{\bibinfo{person}{Nazanin Andalibi}.} \bibinfo{year}{2019}\natexlab{}.
\newblock \showarticletitle{What Happens After Disclosing Stigmatized Experiences on Identified Social Media: Individual, Dyadic, and Social/Network Outcomes}. In \bibinfo{booktitle}{\emph{Proceedings of the 2019 CHI Conference on Human Factors in Computing Systems}} (Glasgow, Scotland Uk) \emph{(\bibinfo{series}{CHI '19})}. \bibinfo{publisher}{Association for Computing Machinery}, \bibinfo{address}{New York, NY, USA}, \bibinfo{pages}{1â€“15}.
\newblock
\showISBNx{9781450359702}
\href{https://doi.org/10.1145/3290605.3300367}{doi:\nolinkurl{10.1145/3290605.3300367}}


\bibitem[Andalibi and Forte(2018)]%
        {andalibi_responding_2018}
\bibfield{author}{\bibinfo{person}{Nazanin Andalibi} {and} \bibinfo{person}{Andrea Forte}.} \bibinfo{year}{2018}\natexlab{}.
\newblock \showarticletitle{Responding to {Sensitive} {Disclosures} on {Social} {Media}: {A} {Decision}-{Making} {Framework}}.
\newblock \bibinfo{journal}{\emph{ACM Transactions on Computer-Human Interaction (TOCHI)}} \bibinfo{volume}{25}, \bibinfo{number}{6} (\bibinfo{year}{2018}), \bibinfo{pages}{31}.
\newblock


\bibitem[Andalibi et~al\mbox{.}(2016)]%
        {andalibi_understanding_2016}
\bibfield{author}{\bibinfo{person}{Nazanin Andalibi}, \bibinfo{person}{Oliver~L. Haimson}, \bibinfo{person}{Munmun De~Choudhury}, {and} \bibinfo{person}{Andrea Forte}.} \bibinfo{year}{2016}\natexlab{}.
\newblock \showarticletitle{Understanding {Social} {Media} {Disclosures} of {Sexual} {Abuse} {Through} the {Lenses} of {Support} {Seeking} and {Anonymity}}. In \bibinfo{booktitle}{\emph{Proceedings of the 2016 {CHI} {Conference} on {Human} {Factors} in {Computing} {Systems}}} \emph{(\bibinfo{series}{{CHI} '16})}. \bibinfo{publisher}{ACM}, \bibinfo{address}{New York, NY, USA}, \bibinfo{pages}{3906--3918}.
\newblock
\showISBNx{978-1-4503-3362-7}
\href{https://doi.org/10.1145/2858036.2858096}{doi:\nolinkurl{10.1145/2858036.2858096}}


\bibitem[Baillargeon et~al\mbox{.}(2025)]%
        {baillargeon2025social}
\bibfield{author}{\bibinfo{person}{Philip Baillargeon}, \bibinfo{person}{Jina Yoon}, {and} \bibinfo{person}{Amy Zhang}.} \bibinfo{year}{2025}\natexlab{}.
\newblock \showarticletitle{Who Puts the "Social" in "Social Computing"?: Using A Neurodiversity Framing to Review Social Computing Research}.
\newblock \bibinfo{journal}{\emph{Proc. ACM Hum.-Comput. Interact.}} \bibinfo{volume}{9}, \bibinfo{number}{2}, Article \bibinfo{articleno}{CSCW208} (\bibinfo{date}{May} \bibinfo{year}{2025}), \bibinfo{numpages}{44}~pages.
\newblock
\href{https://doi.org/10.1145/3711106}{doi:\nolinkurl{10.1145/3711106}}


\bibitem[Baumer et~al\mbox{.}(2013)]%
        {baumer_et_al_13}
\bibfield{author}{\bibinfo{person}{Eric~P.S. Baumer}, \bibinfo{person}{Phil Adams}, \bibinfo{person}{Vera~D. Khovanskaya}, \bibinfo{person}{Tony~C. Liao}, \bibinfo{person}{Madeline~E. Smith}, \bibinfo{person}{Victoria Schwanda~Sosik}, {and} \bibinfo{person}{Kaiton Williams}.} \bibinfo{year}{2013}\natexlab{}.
\newblock \showarticletitle{Limiting, leaving, and (re)lapsing: an exploration of facebook non-use practices and experiences}. In \bibinfo{booktitle}{\emph{Proceedings of the SIGCHI Conference on Human Factors in Computing Systems}} (Paris, France) \emph{(\bibinfo{series}{CHI '13})}. \bibinfo{publisher}{Association for Computing Machinery}, \bibinfo{address}{New York, NY, USA}, \bibinfo{pages}{3257–3266}.
\newblock
\showISBNx{9781450318990}
\href{https://doi.org/10.1145/2470654.2466446}{doi:\nolinkurl{10.1145/2470654.2466446}}


\bibitem[Baumer et~al\mbox{.}(2014)]%
        {baumer_et_al_14}
\bibfield{author}{\bibinfo{person}{Eric~P.S. Baumer}, \bibinfo{person}{Morgan~G. Ames}, \bibinfo{person}{Jed~R. Brubaker}, \bibinfo{person}{Jenna Burrell}, {and} \bibinfo{person}{Paul Dourish}.} \bibinfo{year}{2014}\natexlab{}.
\newblock \showarticletitle{Refusing, limiting, departing: why we should study technology non-use}. In \bibinfo{booktitle}{\emph{CHI '14 Extended Abstracts on Human Factors in Computing Systems}} (Toronto, Ontario, Canada) \emph{(\bibinfo{series}{CHI EA '14})}. \bibinfo{publisher}{Association for Computing Machinery}, \bibinfo{address}{New York, NY, USA}, \bibinfo{pages}{65–68}.
\newblock
\showISBNx{9781450324748}
\href{https://doi.org/10.1145/2559206.2559224}{doi:\nolinkurl{10.1145/2559206.2559224}}


\bibitem[Bennett and Rosner(2019)]%
        {bennett2019promise}
\bibfield{author}{\bibinfo{person}{Cynthia~L. Bennett} {and} \bibinfo{person}{Daniela~K. Rosner}.} \bibinfo{year}{2019}\natexlab{}.
\newblock \showarticletitle{The Promise of Empathy: Design, Disability, and Knowing the "Other"}. In \bibinfo{booktitle}{\emph{Proceedings of the 2019 CHI Conference on Human Factors in Computing Systems}} (Glasgow, Scotland Uk) \emph{(\bibinfo{series}{CHI '19})}. \bibinfo{publisher}{Association for Computing Machinery}, \bibinfo{address}{New York, NY, USA}, \bibinfo{pages}{1–13}.
\newblock
\showISBNx{9781450359702}
\href{https://doi.org/10.1145/3290605.3300528}{doi:\nolinkurl{10.1145/3290605.3300528}}


\bibitem[Bhakuni and Abimbola(2021)]%
        {bhakuni_epistemic_2021}
\bibfield{author}{\bibinfo{person}{Himani Bhakuni} {and} \bibinfo{person}{Seye Abimbola}.} \bibinfo{year}{2021}\natexlab{}.
\newblock \showarticletitle{Epistemic injustice in academic global health}.
\newblock \bibinfo{journal}{\emph{The Lancet Global Health}} \bibinfo{volume}{9}, \bibinfo{number}{10} (\bibinfo{date}{Oct.} \bibinfo{year}{2021}), \bibinfo{pages}{e1465--e1470}.
\newblock
\showISSN{2214-109X}
\href{https://doi.org/10.1016/S2214-109X(21)00301-6}{doi:\nolinkurl{10.1016/S2214-109X(21)00301-6}}
\newblock
\shownote{Publisher: Elsevier}.


\bibitem[Branham and Kane(2015)]%
        {branham2015invisible}
\bibfield{author}{\bibinfo{person}{Stacy~M. Branham} {and} \bibinfo{person}{Shaun~K. Kane}.} \bibinfo{year}{2015}\natexlab{}.
\newblock \showarticletitle{The Invisible Work of Accessibility: How Blind Employees Manage Accessibility in Mixed-Ability Workplaces}. In \bibinfo{booktitle}{\emph{Proceedings of the 17th International ACM SIGACCESS Conference on Computers \& Accessibility}} (Lisbon, Portugal) \emph{(\bibinfo{series}{ASSETS '15})}. \bibinfo{publisher}{Association for Computing Machinery}, \bibinfo{address}{New York, NY, USA}, \bibinfo{pages}{163–171}.
\newblock
\showISBNx{9781450334006}
\href{https://doi.org/10.1145/2700648.2809864}{doi:\nolinkurl{10.1145/2700648.2809864}}


\bibitem[Carel and Kidd(2014)]%
        {carel2014epistemic}
\bibfield{author}{\bibinfo{person}{Havi Carel} {and} \bibinfo{person}{Ian~James Kidd}.} \bibinfo{year}{2014}\natexlab{}.
\newblock \showarticletitle{Epistemic injustice in healthcare: a philosophial analysis}.
\newblock \bibinfo{journal}{\emph{Medicine, Health Care and Philosophy}} \bibinfo{volume}{17}, \bibinfo{number}{4} (\bibinfo{year}{2014}), \bibinfo{pages}{529--540}.
\newblock


\bibitem[Carel and Kidd(2017)]%
        {carel2017epistemic}
\bibfield{author}{\bibinfo{person}{Havi Carel} {and} \bibinfo{person}{Ian~James Kidd}.} \bibinfo{year}{2017}\natexlab{}.
\newblock \showarticletitle{Epistemic injustice in medicine and healthcare}.
\newblock In \bibinfo{booktitle}{\emph{The Routledge handbook of epistemic injustice}}. \bibinfo{publisher}{Routledge}, \bibinfo{pages}{336--346}.
\newblock


\bibitem[Cavanna and Seri(2015)]%
        {cavanna2015misophonia}
\bibfield{author}{\bibinfo{person}{Andrea~E Cavanna} {and} \bibinfo{person}{Stefano Seri}.} \bibinfo{year}{2015}\natexlab{}.
\newblock \showarticletitle{Misophonia: current perspectives}.
\newblock \bibinfo{journal}{\emph{Neuropsychiatric disease and treatment}} (\bibinfo{year}{2015}), \bibinfo{pages}{2117--2123}.
\newblock


\bibitem[Chen et~al\mbox{.}(2022)]%
        {chen_trauma-informed_2022}
\bibfield{author}{\bibinfo{person}{Janet~X. Chen}, \bibinfo{person}{Allison McDonald}, \bibinfo{person}{Yixin Zou}, \bibinfo{person}{Emily Tseng}, \bibinfo{person}{Kevin~A Roundy}, \bibinfo{person}{Acar Tamersoy}, \bibinfo{person}{Florian Schaub}, \bibinfo{person}{Thomas Ristenpart}, {and} \bibinfo{person}{Nicola Dell}.} \bibinfo{year}{2022}\natexlab{}.
\newblock \showarticletitle{Trauma-Informed Computing: Towards Safer Technology Experiences for All}. In \bibinfo{booktitle}{\emph{Proceedings of the 2022 CHI Conference on Human Factors in Computing Systems}} (New Orleans, LA, USA) \emph{(\bibinfo{series}{CHI '22})}. \bibinfo{publisher}{Association for Computing Machinery}, \bibinfo{address}{New York, NY, USA}, Article \bibinfo{articleno}{544}, \bibinfo{numpages}{20}~pages.
\newblock
\showISBNx{9781450391573}
\href{https://doi.org/10.1145/3491102.3517475}{doi:\nolinkurl{10.1145/3491102.3517475}}


\bibitem[Das et~al\mbox{.}(2021)]%
        {das_et_al_21}
\bibfield{author}{\bibinfo{person}{Maitraye Das}, \bibinfo{person}{John Tang}, \bibinfo{person}{Kathryn~E. Ringland}, {and} \bibinfo{person}{Anne~Marie Piper}.} \bibinfo{year}{2021}\natexlab{}.
\newblock \showarticletitle{Towards Accessible Remote Work: Understanding Work-from-Home Practices of Neurodivergent Professionals}.
\newblock \bibinfo{journal}{\emph{Proc. ACM Hum.-Comput. Interact.}} \bibinfo{volume}{5}, \bibinfo{number}{CSCW1}, Article \bibinfo{articleno}{183} (\bibinfo{date}{apr} \bibinfo{year}{2021}), \bibinfo{numpages}{30}~pages.
\newblock
\href{https://doi.org/10.1145/3449282}{doi:\nolinkurl{10.1145/3449282}}


\bibitem[Fricker(2007)]%
        {fricker_epistemic_2007}
\bibfield{author}{\bibinfo{person}{Miranda Fricker}.} \bibinfo{year}{2007}\natexlab{}.
\newblock \bibinfo{booktitle}{\emph{Epistemic {Injustice}: {Power} and the {Ethics} of {Knowing}}}.
\newblock \bibinfo{publisher}{Clarendon Press}.
\newblock
\showISBNx{978-0-19-823790-7}
\newblock
\shownote{Google-Books-ID: lncSDAAAQBAJ}.


\bibitem[Gorm and Shklovski(2019)]%
        {gorm2019episodic}
\bibfield{author}{\bibinfo{person}{Nanna Gorm} {and} \bibinfo{person}{Irina Shklovski}.} \bibinfo{year}{2019}\natexlab{}.
\newblock \showarticletitle{Episodic use: Practices of care in self-tracking}.
\newblock \bibinfo{journal}{\emph{New Media \& Society}} \bibinfo{volume}{21}, \bibinfo{number}{11-12} (\bibinfo{year}{2019}), \bibinfo{pages}{2505--2521}.
\newblock


\bibitem[Hofmann et~al\mbox{.}(2020)]%
        {hofmann2020living}
\bibfield{author}{\bibinfo{person}{Megan Hofmann}, \bibinfo{person}{Devva Kasnitz}, \bibinfo{person}{Jennifer Mankoff}, {and} \bibinfo{person}{Cynthia~L Bennett}.} \bibinfo{year}{2020}\natexlab{}.
\newblock \showarticletitle{Living Disability Theory: Reflections on Access, Research, and Design}. In \bibinfo{booktitle}{\emph{Proceedings of the 22nd International ACM SIGACCESS Conference on Computers and Accessibility}} (Virtual Event, Greece) \emph{(\bibinfo{series}{ASSETS '20})}. \bibinfo{publisher}{Association for Computing Machinery}, \bibinfo{address}{New York, NY, USA}, Article \bibinfo{articleno}{4}, \bibinfo{numpages}{13}~pages.
\newblock
\showISBNx{9781450371032}
\href{https://doi.org/10.1145/3373625.3416996}{doi:\nolinkurl{10.1145/3373625.3416996}}


\bibitem[Jaswal et~al\mbox{.}(2021)]%
        {jaswal2021misokinesia}
\bibfield{author}{\bibinfo{person}{Sumeet~M Jaswal}, \bibinfo{person}{Andreas~KF De~Bleser}, {and} \bibinfo{person}{Todd~C Handy}.} \bibinfo{year}{2021}\natexlab{}.
\newblock \showarticletitle{Misokinesia is a sensitivity to seeing others fidget that is prevalent in the general population}.
\newblock \bibinfo{journal}{\emph{Scientific reports}} \bibinfo{volume}{11}, \bibinfo{number}{1} (\bibinfo{year}{2021}), \bibinfo{pages}{17204}.
\newblock


\bibitem[Jiang et~al\mbox{.}(2025)]%
        {jiang2025shifting}
\bibfield{author}{\bibinfo{person}{Lucy Jiang}, \bibinfo{person}{Woojin Ko}, \bibinfo{person}{Shirley Yuan}, \bibinfo{person}{Tanisha Shende}, {and} \bibinfo{person}{Shiri Azenkot}.} \bibinfo{year}{2025}\natexlab{}.
\newblock \showarticletitle{Shifting the Focus: Exploring Video Accessibility Strategies and Challenges for People with ADHD}. In \bibinfo{booktitle}{\emph{Proceedings of the 2025 CHI Conference on Human Factors in Computing Systems}} \emph{(\bibinfo{series}{CHI '25})}. \bibinfo{publisher}{Association for Computing Machinery}, \bibinfo{address}{New York, NY, USA}, Article \bibinfo{articleno}{561}, \bibinfo{numpages}{16}~pages.
\newblock
\showISBNx{9798400713941}
\href{https://doi.org/10.1145/3706598.3713637}{doi:\nolinkurl{10.1145/3706598.3713637}}


\bibitem[Kidd and Carel(2017)]%
        {kidd2017epistemic}
\bibfield{author}{\bibinfo{person}{Ian~James Kidd} {and} \bibinfo{person}{Havi Carel}.} \bibinfo{year}{2017}\natexlab{}.
\newblock \showarticletitle{Epistemic injustice and illness}.
\newblock \bibinfo{journal}{\emph{Journal of applied philosophy}} \bibinfo{volume}{34}, \bibinfo{number}{2} (\bibinfo{year}{2017}), \bibinfo{pages}{172--190}.
\newblock


\bibitem[Li et~al\mbox{.}(2024)]%
        {li2024codesigning}
\bibfield{author}{\bibinfo{person}{Jingjin Li}, \bibinfo{person}{Shaomei Wu}, {and} \bibinfo{person}{Gilly Leshed}.} \bibinfo{year}{2024}\natexlab{}.
\newblock \showarticletitle{Re-envisioning Remote Meetings: Co-designing Inclusive and Empowering Videoconferencing with People Who Stutter}. In \bibinfo{booktitle}{\emph{Proceedings of the 2024 ACM Designing Interactive Systems Conference}} (Copenhagen, Denmark) \emph{(\bibinfo{series}{DIS '24})}. \bibinfo{publisher}{Association for Computing Machinery}, \bibinfo{address}{New York, NY, USA}, \bibinfo{pages}{1926–1941}.
\newblock
\showISBNx{9798400705830}
\href{https://doi.org/10.1145/3643834.3661533}{doi:\nolinkurl{10.1145/3643834.3661533}}


\bibitem[Mack et~al\mbox{.}(2021)]%
        {mack2021what}
\bibfield{author}{\bibinfo{person}{Kelly Mack}, \bibinfo{person}{Emma McDonnell}, \bibinfo{person}{Dhruv Jain}, \bibinfo{person}{Lucy Lu~Wang}, \bibinfo{person}{Jon E.~Froehlich}, {and} \bibinfo{person}{Leah Findlater}.} \bibinfo{year}{2021}\natexlab{}.
\newblock \showarticletitle{What Do We Mean by “Accessibility Research”? A Literature Survey of Accessibility Papers in CHI and ASSETS from 1994 to 2019}. In \bibinfo{booktitle}{\emph{Proceedings of the 2021 CHI Conference on Human Factors in Computing Systems}} (Yokohama, Japan) \emph{(\bibinfo{series}{CHI '21})}. \bibinfo{publisher}{Association for Computing Machinery}, \bibinfo{address}{New York, NY, USA}, Article \bibinfo{articleno}{371}, \bibinfo{numpages}{18}~pages.
\newblock
\showISBNx{9781450380966}
\href{https://doi.org/10.1145/3411764.3445412}{doi:\nolinkurl{10.1145/3411764.3445412}}


\bibitem[Mankoff et~al\mbox{.}(2010)]%
        {mankoff_disability_2010}
\bibfield{author}{\bibinfo{person}{Jennifer Mankoff}, \bibinfo{person}{Gillian~R. Hayes}, {and} \bibinfo{person}{Devva Kasnitz}.} \bibinfo{year}{2010}\natexlab{}.
\newblock \showarticletitle{Disability studies as a source of critical inquiry for the field of assistive technology}. In \bibinfo{booktitle}{\emph{Proceedings of the 12th international {ACM} {SIGACCESS} conference on {Computers} and accessibility}} \emph{(\bibinfo{series}{{ASSETS} '10})}. \bibinfo{publisher}{Association for Computing Machinery}, \bibinfo{address}{New York, NY, USA}, \bibinfo{pages}{3--10}.
\newblock
\showISBNx{978-1-60558-881-0}
\href{https://doi.org/10.1145/1878803.1878807}{doi:\nolinkurl{10.1145/1878803.1878807}}


\bibitem[McKinnon(2016)]%
        {mckinnon2016epistemic}
\bibfield{author}{\bibinfo{person}{Rachel McKinnon}.} \bibinfo{year}{2016}\natexlab{}.
\newblock \showarticletitle{Epistemic injustice}.
\newblock \bibinfo{journal}{\emph{Philosophy Compass}} \bibinfo{volume}{11}, \bibinfo{number}{8} (\bibinfo{year}{2016}), \bibinfo{pages}{437--446}.
\newblock


\bibitem[Mohamed(2024)]%
        {mohamed2024debilitating}
\bibfield{author}{\bibinfo{person}{Kharnita Mohamed}.} \bibinfo{year}{2024}\natexlab{}.
\newblock \showarticletitle{Debilitating Research: Scholarship of the Obvious and Epistemic Trauma}.
\newblock \bibinfo{journal}{\emph{African Studies}} \bibinfo{volume}{83}, \bibinfo{number}{2-3} (\bibinfo{year}{2024}), \bibinfo{pages}{134--151}.
\newblock


\bibitem[Nevsky et~al\mbox{.}(2025)]%
        {nevskyetal25}
\bibfield{author}{\bibinfo{person}{Alexandre Nevsky}, \bibinfo{person}{Filip Bircanin}, \bibinfo{person}{Elena Simperl}, \bibinfo{person}{Madeline~N Cruice}, {and} \bibinfo{person}{Timothy Neate}.} \bibinfo{year}{2025}\natexlab{}.
\newblock \showarticletitle{To Each Their Own: Exploring Highly Personalised Audiovisual Media Accessibility Interventions with People with Aphasia}. In \bibinfo{booktitle}{\emph{Proceedings of the 2025 ACM Designing Interactive Systems Conference}} \emph{(\bibinfo{series}{DIS '25})}. \bibinfo{publisher}{Association for Computing Machinery}, \bibinfo{address}{New York, NY, USA}, \bibinfo{pages}{1826–1843}.
\newblock
\showISBNx{9798400714856}
\href{https://doi.org/10.1145/3715336.3735771}{doi:\nolinkurl{10.1145/3715336.3735771}}


\bibitem[Newbigging and Ridley(2018)]%
        {newbigging2018epistemic}
\bibfield{author}{\bibinfo{person}{Karen Newbigging} {and} \bibinfo{person}{Julie Ridley}.} \bibinfo{year}{2018}\natexlab{}.
\newblock \showarticletitle{Epistemic struggles: The role of advocacy in promoting epistemic justice and rights in mental health}.
\newblock \bibinfo{journal}{\emph{Social Science \& Medicine}}  \bibinfo{volume}{219} (\bibinfo{year}{2018}), \bibinfo{pages}{36--44}.
\newblock


\bibitem[Potgieter et~al\mbox{.}(2019)]%
        {potgieter2019misophonia}
\bibfield{author}{\bibinfo{person}{Iskra Potgieter}, \bibinfo{person}{Carol MacDonald}, \bibinfo{person}{Lucy Partridge}, \bibinfo{person}{Rilana Cima}, \bibinfo{person}{Jacqueline Sheldrake}, {and} \bibinfo{person}{Derek~J Hoare}.} \bibinfo{year}{2019}\natexlab{}.
\newblock \showarticletitle{Misophonia: A scoping review of research}.
\newblock \bibinfo{journal}{\emph{Journal of clinical psychology}} \bibinfo{volume}{75}, \bibinfo{number}{7} (\bibinfo{year}{2019}), \bibinfo{pages}{1203--1218}.
\newblock


\bibitem[Poulsen et~al\mbox{.}(2024)]%
        {poulsen2024auditory}
\bibfield{author}{\bibinfo{person}{Rebecca Poulsen}, \bibinfo{person}{Z Williams}, \bibinfo{person}{Patrick Dwyer}, \bibinfo{person}{E Pellicano}, \bibinfo{person}{Paul~F Sowman}, {and} \bibinfo{person}{David McAlpine}.} \bibinfo{year}{2024}\natexlab{}.
\newblock \showarticletitle{How auditory processing influences the autistic profile: A review}.
\newblock \bibinfo{journal}{\emph{Autism Research}} \bibinfo{volume}{17}, \bibinfo{number}{12} (\bibinfo{year}{2024}), \bibinfo{pages}{2452--2470}.
\newblock


\bibitem[Race et~al\mbox{.}(2021)]%
        {race_et_al_21}
\bibfield{author}{\bibinfo{person}{Lauren Race}, \bibinfo{person}{Amber James}, \bibinfo{person}{Andrew Hayward}, \bibinfo{person}{Kia El-Amin}, \bibinfo{person}{Maya~Gold Patterson}, {and} \bibinfo{person}{Theresa Mershon}.} \bibinfo{year}{2021}\natexlab{}.
\newblock \showarticletitle{Designing Sensory and Social Tools for Neurodivergent Individuals in Social Media Environments}. In \bibinfo{booktitle}{\emph{Proceedings of the 23rd International ACM SIGACCESS Conference on Computers and Accessibility}} (Virtual Event, USA) \emph{(\bibinfo{series}{ASSETS '21})}. \bibinfo{publisher}{Association for Computing Machinery}, \bibinfo{address}{New York, NY, USA}, Article \bibinfo{articleno}{61}, \bibinfo{numpages}{5}~pages.
\newblock
\showISBNx{9781450383066}
\href{https://doi.org/10.1145/3441852.3476546}{doi:\nolinkurl{10.1145/3441852.3476546}}


\bibitem[Randazzo and Ammari(2023)]%
        {randazzo_ammari_23}
\bibfield{author}{\bibinfo{person}{Casey Randazzo} {and} \bibinfo{person}{Tawfiq Ammari}.} \bibinfo{year}{2023}\natexlab{}.
\newblock \showarticletitle{“If Someone Downvoted My Posts—That’d Be the End of the World”: Designing Safer Online Spaces for Trauma Survivors}. In \bibinfo{booktitle}{\emph{Proceedings of the 2023 CHI Conference on Human Factors in Computing Systems}} (Hamburg, Germany) \emph{(\bibinfo{series}{CHI '23})}. \bibinfo{publisher}{Association for Computing Machinery}, \bibinfo{address}{New York, NY, USA}, Article \bibinfo{articleno}{481}, \bibinfo{numpages}{18}~pages.
\newblock
\showISBNx{9781450394215}
\href{https://doi.org/10.1145/3544548.3581453}{doi:\nolinkurl{10.1145/3544548.3581453}}


\bibitem[Randazzo and Ammari(2025)]%
        {randazzo_kintsugi_25}
\bibfield{author}{\bibinfo{person}{Casey Randazzo} {and} \bibinfo{person}{Tawfiq Ammari}.} \bibinfo{year}{2025}\natexlab{}.
\newblock \showarticletitle{Kintsugi-Inspired Design: Communicatively Reconstructing Identities Online After Trauma}.
\newblock \bibinfo{journal}{\emph{Proc. ACM Hum.-Comput. Interact.}} \bibinfo{volume}{9}, \bibinfo{number}{7}, Article \bibinfo{articleno}{CSCW326} (\bibinfo{date}{Oct.} \bibinfo{year}{2025}), \bibinfo{numpages}{31}~pages.
\newblock
\href{https://doi.org/10.1145/3757507}{doi:\nolinkurl{10.1145/3757507}}


\bibitem[Randazzo et~al\mbox{.}(2023)]%
        {randazzo_workshop_23}
\bibfield{author}{\bibinfo{person}{Casey Randazzo}, \bibinfo{person}{Carol~F. Scott}, \bibinfo{person}{Rosanna Bellini}, \bibinfo{person}{Tawfiq Ammari}, \bibinfo{person}{Michael~Ann Devito}, \bibinfo{person}{Bryan Semaan}, {and} \bibinfo{person}{Nazanin Andalibi}.} \bibinfo{year}{2023}\natexlab{}.
\newblock \showarticletitle{Trauma-Informed Design: A Collaborative Approach to Building Safer Online Spaces}. In \bibinfo{booktitle}{\emph{Companion Publication of the 2023 Conference on Computer Supported Cooperative Work and Social Computing}} (Minneapolis, MN, USA) \emph{(\bibinfo{series}{CSCW '23 Companion})}. \bibinfo{publisher}{Association for Computing Machinery}, \bibinfo{address}{New York, NY, USA}, \bibinfo{pages}{470–475}.
\newblock
\showISBNx{9798400701290}
\href{https://doi.org/10.1145/3584931.3611277}{doi:\nolinkurl{10.1145/3584931.3611277}}


\bibitem[Sabinson(2025a)]%
        {sabinson2024pictorial}
\bibfield{author}{\bibinfo{person}{Elena Sabinson}.} \bibinfo{year}{2025}\natexlab{a}.
\newblock \showarticletitle{Blowfish Band: A Wearable Inflatable Fidget for Self-Stimulatory (Stim) Behaviors}. In \bibinfo{booktitle}{\emph{Proceedings of the 2025 ACM Designing Interactive Systems Conference}} \emph{(\bibinfo{series}{DIS '25})}. \bibinfo{publisher}{Association for Computing Machinery}, \bibinfo{address}{New York, NY, USA}, \bibinfo{pages}{1–0}.
\newblock
\showISBNx{9798400714856}
\href{https://doi.org/10.1145/3715336.3735442}{doi:\nolinkurl{10.1145/3715336.3735442}}


\bibitem[Sabinson(2025b)]%
        {sabinson2025blowfish}
\bibfield{author}{\bibinfo{person}{Elena Sabinson}.} \bibinfo{year}{2025}\natexlab{b}.
\newblock \showarticletitle{Blowfish Band: A Wearable Inflatable Fidget for Self-Stimulatory (Stim) Behaviors}. In \bibinfo{booktitle}{\emph{Proceedings of the 2025 ACM Designing Interactive Systems Conference}} \emph{(\bibinfo{series}{DIS '25})}. \bibinfo{publisher}{Association for Computing Machinery}, \bibinfo{address}{New York, NY, USA}, \bibinfo{pages}{1–0}.
\newblock
\showISBNx{9798400714856}
\href{https://doi.org/10.1145/3715336.3735442}{doi:\nolinkurl{10.1145/3715336.3735442}}


\bibitem[Salovaara et~al\mbox{.}(2011)]%
        {salovaara2011everyday}
\bibfield{author}{\bibinfo{person}{Antti Salovaara}, \bibinfo{person}{Sacha Helfenstein}, {and} \bibinfo{person}{Antti Oulasvirta}.} \bibinfo{year}{2011}\natexlab{}.
\newblock \showarticletitle{Everyday appropriations of information technology: A study of creative uses of digital cameras}.
\newblock \bibinfo{journal}{\emph{Journal of the American Society for Information Science and Technology}} \bibinfo{volume}{62}, \bibinfo{number}{12} (\bibinfo{year}{2011}), \bibinfo{pages}{2347--2363}.
\newblock


\bibitem[Schr{\"o}der et~al\mbox{.}(2013)]%
        {schroder2013misophonia}
\bibfield{author}{\bibinfo{person}{Arjan Schr{\"o}der}, \bibinfo{person}{Nienke Vulink}, {and} \bibinfo{person}{Damiaan Denys}.} \bibinfo{year}{2013}\natexlab{}.
\newblock \showarticletitle{Misophonia: diagnostic criteria for a new psychiatric disorder}.
\newblock \bibinfo{journal}{\emph{PloS one}} \bibinfo{volume}{8}, \bibinfo{number}{1} (\bibinfo{year}{2013}), \bibinfo{pages}{e54706}.
\newblock


\bibitem[Scott et~al\mbox{.}(2023)]%
        {Scott2023}
\bibfield{author}{\bibinfo{person}{Carol~F Scott}, \bibinfo{person}{Gabriela Marcu}, \bibinfo{person}{Riana~Elyse Anderson}, \bibinfo{person}{Mark~W Newman}, {and} \bibinfo{person}{Sarita Schoenebeck}.} \bibinfo{year}{2023}\natexlab{}.
\newblock \showarticletitle{Trauma-Informed Social Media: Towards Solutions for Reducing and Healing Online Harm}. In \bibinfo{booktitle}{\emph{Proceedings of the 2023 CHI Conference on Human Factors in Computing Systems}} (Hamburg, Germany) \emph{(\bibinfo{series}{CHI '23})}. \bibinfo{publisher}{Association for Computing Machinery}, \bibinfo{address}{New York, NY, USA}, Article \bibinfo{articleno}{341}, \bibinfo{numpages}{20}~pages.
\newblock
\showISBNx{9781450394215}
\href{https://doi.org/10.1145/3544548.3581512}{doi:\nolinkurl{10.1145/3544548.3581512}}


\bibitem[{soQuiet Misophonia Advocacy}(2024a)]%
        {soquiet_mrn}
\bibfield{author}{\bibinfo{person}{{soQuiet Misophonia Advocacy}}.} \bibinfo{year}{2024}\natexlab{a}.
\newblock \bibinfo{title}{{Misophonia Research Network}}.
\newblock \bibinfo{howpublished}{\url{https://www.soquiet.org/mrn}}.
\newblock
\newblock
\shownote{Accessed: 2026-05-10}.


\bibitem[{soQuiet Misophonia Advocacy}(2024b)]%
        {soquiet_ammari}
\bibfield{author}{\bibinfo{person}{{soQuiet Misophonia Advocacy}}.} \bibinfo{year}{2024}\natexlab{b}.
\newblock \bibinfo{title}{{Tawfiq Ammari, PhD}}.
\newblock \bibinfo{howpublished}{\url{https://www.soquiet.org/tawfiq-ammari-phd}}.
\newblock
\newblock
\shownote{Accessed: 2026-05-10. Researcher profile, soQuiet Misophonia Research Network.}.


\bibitem[S{\o}rensen(2006)]%
        {sorensen2006domestication}
\bibfield{author}{\bibinfo{person}{Knut~H S{\o}rensen}.} \bibinfo{year}{2006}\natexlab{}.
\newblock \showarticletitle{Domestication: the enactment of technology}.
\newblock \bibinfo{journal}{\emph{Domestication of media and technology}}  \bibinfo{volume}{46} (\bibinfo{year}{2006}).
\newblock


\bibitem[Spiel et~al\mbox{.}(2020)]%
        {spiel2020nothing}
\bibfield{author}{\bibinfo{person}{Katta Spiel}, \bibinfo{person}{Kathrin Gerling}, \bibinfo{person}{Cynthia~L. Bennett}, \bibinfo{person}{Emeline Brul\'{e}}, \bibinfo{person}{Rua~M. Williams}, \bibinfo{person}{Jennifer Rode}, {and} \bibinfo{person}{Jennifer Mankoff}.} \bibinfo{year}{2020}\natexlab{}.
\newblock \showarticletitle{Nothing About Us Without Us: Investigating the Role of Critical Disability Studies in HCI}. In \bibinfo{booktitle}{\emph{Extended Abstracts of the 2020 CHI Conference on Human Factors in Computing Systems}} (Honolulu, HI, USA) \emph{(\bibinfo{series}{CHI EA '20})}. \bibinfo{publisher}{Association for Computing Machinery}, \bibinfo{address}{New York, NY, USA}, \bibinfo{pages}{1–8}.
\newblock
\showISBNx{9781450368193}
\href{https://doi.org/10.1145/3334480.3375150}{doi:\nolinkurl{10.1145/3334480.3375150}}


\bibitem[Stinnett(2018)]%
        {stinnett2018trauma}
\bibfield{author}{\bibinfo{person}{Gina Stinnett}.} \bibinfo{year}{2018}\natexlab{}.
\newblock \bibinfo{booktitle}{\emph{Trauma and the Credibility Economy: An Analysis of Epistemic Violence and Its Traumatic Functions}}.
\newblock \bibinfo{publisher}{Illinois State University}.
\newblock


\bibitem[Swedo et~al\mbox{.}(2022)]%
        {swedo2022consensus}
\bibfield{author}{\bibinfo{person}{Susan~E Swedo}, \bibinfo{person}{David~M Baguley}, \bibinfo{person}{Damiaan Denys}, \bibinfo{person}{Laura~J Dixon}, \bibinfo{person}{Mercede Erfanian}, \bibinfo{person}{Alessandra Fioretti}, \bibinfo{person}{Pawel~J Jastreboff}, \bibinfo{person}{Sukhbinder Kumar}, \bibinfo{person}{M~Zachary Rosenthal}, \bibinfo{person}{Romke Rouw}, {et~al\mbox{.}}} \bibinfo{year}{2022}\natexlab{}.
\newblock \showarticletitle{Consensus definition of misophonia: a delphi study}.
\newblock \bibinfo{journal}{\emph{Frontiers in Neuroscience}}  \bibinfo{volume}{16} (\bibinfo{year}{2022}), \bibinfo{pages}{841816}.
\newblock


\bibitem[Taylor(2017)]%
        {taylor2017misophonia}
\bibfield{author}{\bibinfo{person}{Steven Taylor}.} \bibinfo{year}{2017}\natexlab{}.
\newblock \showarticletitle{Misophonia: A new mental disorder?}
\newblock \bibinfo{journal}{\emph{Medical Hypotheses}}  \bibinfo{volume}{103} (\bibinfo{year}{2017}), \bibinfo{pages}{109--117}.
\newblock


\bibitem[Zolyomi and Snyder(2021)]%
        {zolyomi2021social}
\bibfield{author}{\bibinfo{person}{Annuska Zolyomi} {and} \bibinfo{person}{Jaime Snyder}.} \bibinfo{year}{2021}\natexlab{}.
\newblock \showarticletitle{Social-Emotional-Sensory Design Map for Affective Computing Informed by Neurodivergent Experiences}.
\newblock \bibinfo{journal}{\emph{Proc. ACM Hum.-Comput. Interact.}} \bibinfo{volume}{5}, \bibinfo{number}{CSCW1}, Article \bibinfo{articleno}{77} (\bibinfo{date}{April} \bibinfo{year}{2021}), \bibinfo{numpages}{37}~pages.
\newblock
\href{https://doi.org/10.1145/3449151}{doi:\nolinkurl{10.1145/3449151}}


\end{thebibliography}

\end{document}